\documentclass[11pt,english,twoside]{article}

\usepackage{amssymb}
\usepackage{epsfig}
\usepackage{graphics}
\usepackage{graphicx}
\usepackage{amsfonts}
\usepackage{amsmath}
\usepackage{pdflscape}
\usepackage{latexsym}
\usepackage{authblk}

\usepackage[latin1]{inputenc}
\usepackage[english]{babel}

\usepackage{booktabs}
\usepackage{ctable}
\usepackage{subfigure}
\usepackage{float}
\usepackage{layout}
\usepackage{fancyhdr}
\usepackage{color}
\usepackage{eurosym}   
\usepackage{wrapfig}  

\usepackage{xr}
\usepackage{amssymb}
\usepackage{amsmath}
\usepackage{amsfonts}

\input amssym.def
\input amssym.tex

\date{}

\title{The dynamics of the prices of the companies of the STOXX Europe 600 Index through the logit model and neural network}
\author{Federico Mecchia}
\affil{Master in Economics and Finance, Italy}
\author{Marcellino Gaudenzi}
\affil{Dipartimento di Scienze economiche e statistiche, Università di Udine, via Tomadini, 30/A, Italy}

\begin{document}

\maketitle

\begin{abstract}
The aim of the present work is analysing and understanding the dynamics of the prices of companies, depending on whether they are included or excluded from the STOXX Europe 600 Index. For this reason, data regarding the companies of the Index in question was collected and analysed also through the use of logit models and neural networks in order to find the independent variables that affect the changes in prices and thus determine the dynamics over time.
\end{abstract}


\bigskip


\section{Introduction}
Financial indexes today play the role of protagonists in the worldwide economy and, in fact, they are often taken into consideration to acquire an immediate and well-rounded sight about a specific market or about particular trends. In addition to this, because of the important impact of these indexes on the global economy, determining the dynamics of the prices of the stocks of the companies of the various indexes can be of fundamental importance, also with regard to the assessment of other correlated dynamics and the impact on the economic world as a whole.

For these reasons, the focus of this work is understanding the dynamics of the prices of stocks of the companies of the STOXX Europe 600 Index; for this purpose, the logit model and the neural network have been taken into consideration and applied, through the ad hoc code written in Python, to the created database in order to reach the aforementioned objective.

As it was stated earlier, indexes play a key role in today's economy and various researches have been carried out with regard to this topic. For example, researchers have investigated different aspects of the S\&P500 Index, of the NASDAQ Index, of the DOWJONES Index and of the NIKKEI Index. To this regard, in fact, for example, Lee and Hao \cite{LH} have considered the S\&P500 Index in order to study the asymmetric effects of error corrections between oil prices in the U.S.A. and the prices of the aforementioned index, while Lento and Gradojevic \cite{LGrado} focused their attention on understanding the price spillovers of the S\&P500 Index related to the effects on the market of COVID-19. Moreover, Metghalchi et al. \cite{MCH} have explored the trading strategies related to the moving average with regard to the NASDAQ Index, Giannarakis et al. \cite{GPZ} have carried out an analysis of the Dow Jones Sustainability Index and Christiansen and Koldertsova \cite{CK} have taken into account several indexes, such as the NASDAQ Index, in order to explore the role of stock exchanges in corporate governance.

With regard to the STOXX Europe 600 Index, Belhouchet and Chouaibi \cite{BC} took into consideration the STOXX Europe 600 Index to determine the relationship between the integrated reporting quality and the audit committee attributes and Kyriazis \cite{K} analysed the link between COVID-19 deaths, crude oil and gold and the STOXX Europe 600 Index and its sub-indices and STOXX Europe 50. In addition to this, Salvi et al. \cite{STG} studied the relationship between the volume of LBOs and the STOXX Europe 600 Index and De Blasis and Petroni \cite{DP} considered the Index in question to analyse the relationship between oil and renewable energy firms with regard to the COVID-19 pandemic. In addition to this, Langenstein et al. \cite{LUW} took into consideration the companies of the STOXX Europe 600 Index together with the companies of the S\&P500 Index to explore the best strategies that can be enhanced during crisis, Chen et al. \cite{CMPZ} analysed the dynamics of the FTSE 100, S\&P500 Index and STOXX Europe 50 by taking into account different exchange rates and Bonsón and Bednárová \cite{BB} considered some of the companies of the STOXX Europe 600 Index to carry out their research regarding the CSR reporting practices of the companies of the Eurozone. To the authors' knowledge, there are no researches focusing on whether the inclusion or the exclusion of a company from the STOXX Europe 600 Index has an impact on the fluctuations of the prices of the company in question; for this reason, the objective of this work is understanding the dynamics of the prices of companies depending on whether the company considered is a constituent or is not part of the STOXX Europe 600 Index.

For this purpose, several datasets were built by taking into consideration the data that can be found in Bloomberg. These datasets were then analysed through the logit model to determine which are the most significant variables that have an impact on the fluctuations of the prices. The selected variables were then taken into account to build the neural networks for each of the companies taken into account; the levels of accuracy varied according to the company considered, but overall it is possible to state that they were all around the value of 70\%, thus providing a sufficiently satisfactory result.

If compared to the existing literature, the present work is characterized by some distinctive features. Among these, there is the fact that the logistic regression model is used to select the independent variables that are then included in the creation of the neural network and therefore the logit model and the neural networks are not considered to be independent procedures, but part of the whole process. Moreover, another distinguishing aspect of this work is the thorough process of data selection and data analysis that have been carried out. Lastly, another peculiar characteristic of this work is the comparison of the neural networks among various companies of the STOXX Europe 600 Index in order to determine the overall accuracy derived from the application of neural networks to different realities of the Index in question.

\section{Methodology}

As it was stated earlier, the methodology employed in this work in order to analyse the dynamics of the prices of the companies of the STOXX Europe 600 Index involved the use of the logit model and of the neural networks, that have been applied to the ad hoc created database through the code written in Python. More specifically, the logit model was employed in order to determine which are the most significant independent variables that have an impact on the fluctuations of prices of the companies considered; these variables were then taken into account to build the neural networks for each of the companies taken into consideration.

The logit model is a non-linear regression model and it is used in order to determine the probability that a particular event occurs. The dependent variable that is used in this type of model is dichotomic; in this case the dependent variable is represented by the price fluctuation, as it is explained in the following parts of this work. As it was stated earlier, understanding the dynamics of the fluctuations of prices is of fundamental importance, because prices are among the most important characterizing features of data concerning stocks and also because the impact they can potentially have on other aspects of the economy is a key element of various financial scenarios.

In technical terms, for each company taken into account consider the dependent dichotomic variable $Y_i$, which therefore can only be equal to either 0 or 1 and which indicates whether there has been a negative or positive change in price with respect to the prior period, with $i$ ($i=1,\dots,n)$ indicating the specific time instant taken into consideration.
Moreover, consider the vector $x_i=(x_{ij},\dots, x_{im})$ and which includes at the time instant $i$ all the types of data regarding the various characteristics of the company taken into account, indicated by $j$ (with $j = 1,\dots, m)$. The logit model can therefore be defined as:

$$logit(P(Y_i=1|x_{ij} )) = \log \left( \frac{P(Y_i  = 1|x_{ij})}{P(Y_i  = 0|x_{ij})}\right) = \log \left( \frac{P(Y_i  = 1|x_{ij})}{1-P(Y_i  = 1|x_{ij})}\right)= \beta_i$$

Furthermore, the neural networks can be thought of as the artificial transposition of biological systems of neurons that contribute to the learning mechanisms of humans. In other words, as the following image shows (Figure 1), the neurons are structured according to an architecture so that each neuron is connected to the other neurons of the network in question. In addition to this, every neural network has different layers. The first layer is known as the input layer, the last layer is known as the output layer and the layers in the middle are the hidden layers.

\begin{figure}[h!]
	\begin{center}
		\includegraphics[width=0.8\textwidth]{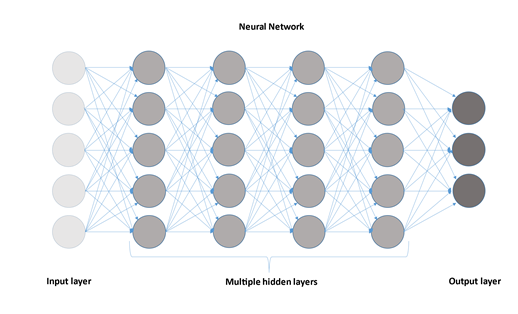}
		\caption{Architecture of neural network}
	    \label{figure1}
		\end{center}
	\end{figure}

Each neuron of a specific layer performs different functions. First it sums up all the values of all the neurons of the previous layer to which it is connected and multiplies these values by the weight, which is located between the previous neuron and the neuron in question; in addition to this, the bias value can also be added. The neuron then uses the activation function which converts the predetermined value to a number between 0 and 1. This process is repeated throughout all the neural network, until the final layer (output layer) is created. Finally, if the final output matches the expected output, it means that the computed parameters are valid and so the process can be carried out with regard to the other inputs; in all other cases the weights are changed and the process is repeated until a satisfactory result is achieved.

With regard to the use of logit models for prediction purposes, various authors have carried out their researches by applying this type of model to different situations. To this regard, for example, Celine et al. \cite{CDD} employed the logit model for the prediction of employability, while Borucka \cite{B} worked with the logistic regression model with regard to the transport services.

In addition to this, other authors have focused their attention on the application of the logistic regression model to the economic and financial sectors. To this regard, for example, Zaini et al. \cite{ZMY} used the logistic regression to analyse the market movements and Ali et al. \cite{AMLH} employed the logistic regression for stock prediction of the PSX. In addition to this, Smita \cite{S} used this type of model for the prediction of the performance of S\&P BSE30 Company and Mironiuc and Robu \cite{MR} used the logit model to predict the performance of stocks on the emerging markets.

With regard to the use of neural networks for price prediction, Mehtab and Sen \cite{MS} employed different tools, including the neural networks, for the prediction of stock prices and Shahvaroughi Farahani and Razavi Hajiagha \cite{SR} studied how neural networks and metaheuristic algorithms can be used for stock prediction. Lastly, Song et al. \cite{SZH} used different types of neural networks to predicts stock prices, Moghaddam et al. \cite{MME} also employed the neural networks to analyse the NASDAQ stock price and also Lachiheb and Gouider \cite{LGouider} worked with neural network for the prices prediction of the TUNINDEX stocks.

\section{Construction of datasets and analysis of data}

As it was stated earlier, Python was employed in order to carry out all the different parts of the analysis and model creation. All the data that was taken into consideration for this work was taken from Bloomberg. More specifically, two sources of data were considered in order to create the final databases that were later involved in the parts inherent to the data analysis.

To this regard, in fact, the first source of data consists of the list of the 600 companies that compose the STOXX Europe 600 Index for each Friday from Friday 4th January 2002 until Friday 31st December 2021. In case the data for the specific Friday taken into consideration was not available, the previous day before the Friday in question for which the data could be collected was taken as reference for that week. This happened in few cases; for example, the data for Friday 09th April 2004 was not available and therefore the data of Thursday 08th April 2004 was considered.

The second source of data, instead, consisted of historical data of each of the companies that were recorded at least once in the aforementioned 20-year period.

In the beginning the first source of data, therefore the lists of the components of the STOXX Europe 600 Index were considered in order to carry out the first phases of the analysis. More specifically, the dynamics regarding the presence or absence of the companies throughout the 20 years period was investigated. To this regard, the number of times a specific company was a constituent of the STOXX Europe 600 Index greatly varies according to the specific company taken into account. In order words, some companies were present just a limited number of times throughout the 20-year period and therefore not for the entire timespan.

The range regarding the number of times a company was a constituent of the STOXX Europe 600 Index was divided into five groups, more specifically each representing one fifth of the range. Each class, as it can be easily inferred from Figure 2, was composed of different number of companies.

\begin{figure}[h!]
	\begin{center}
		\includegraphics[width=0.9\textwidth]{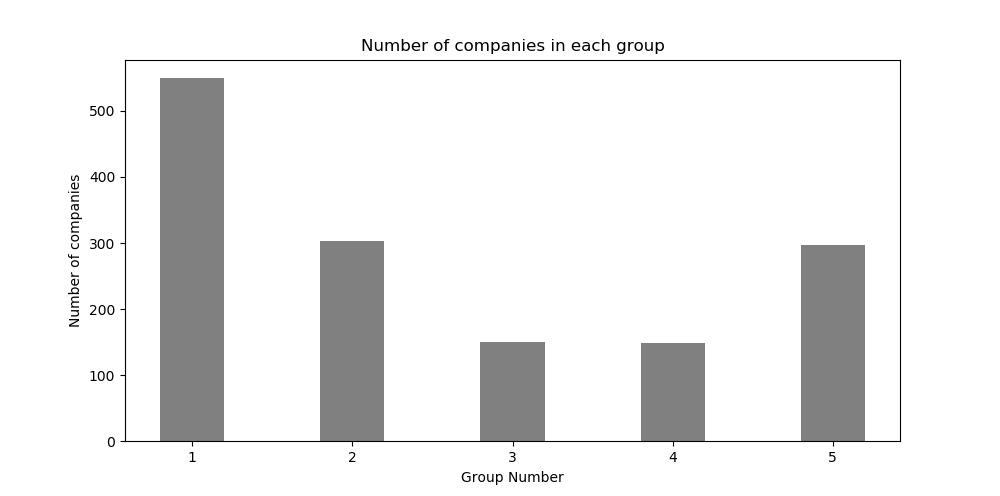}
		\caption{Number of companies in each group}
		\label{figure1}
	\end{center}
\end{figure}

A total of 50 companies, 10 from each of the aforementioned class, were then randomly selected in order to carry out the other parts of the analysis.

As it was previously stated, the second source of data consisted of historical data of each of the 50 companies that were randomly selected. The historical data consisted of financial data, data regarding the member of boards, and other types of data. These files containing the historical data for each company were then merged with the files containing the list of the constituents of the STOXX Europe 600 Index in order to identify when (in order words the specific dates), a particular company was present or absent with regard to the Index in question. For this purpose, a new column was created in each file of each company considered and the column in question was filled through the merge procedure with values equal either to 0 or to 1 (the value 0 indicates that the company was not part of the Index at the date, while the value 1 indicates that was present in the Index at that date).

In addition to this, also another column was added to each dataset. The column in question indicated the dynamics of the prices, in order words whether the price moved up or down with regard to the previous observation. To this regard, the value 1 was assigned whenever there was a positive change in the price and the value 0 was used to fill the column whenever there was a decrease in the price considered.

Moreover, some potential independent variables were considered with a 7-days lag; in other words the values of the previous week of some variables were taken into account and this was done to understand how the data of the previous Friday (or of the previous week) could have an impact on the dependent variable seven days later.

Through the selection of specific companies, it was possible to observe that data regarding the 20-year timespan in some cases was lacking and therefore, depending on the entity of the missing data, some variables were removed. In addition to this, the timespan for each company was designed ad hoc on the basis of the data available; this fact inevitably caused the reduction of the total timespan considered to shorter periods of time depending on the company considered.

The following part of the analysis then consisted in normalizing the various aforementioned datasets. To this regard, the following formula was used in order to normalize each data of the datasets:
$$x_{\hbox{normalized}} = \frac{x-\min_{}{x}}{\max_{}{x} -\min_{}{x}}$$
Therefore, first the minimum and maximum values of each column were computed and then the minimum value was subtracted from the specific data taken into consideration and the result was then divided by the range, defined as the difference between maximum and minimum values. In this way, all the data of the various datasets were converted in values ranging from 0 to 1. Moreover, the missing data was replaced by the mean of the column considered (in other words the mean of the available data for the specific variable taken into account).

The different datasets were then taken into consideration with regard to the logistic models. The logistic models were thus built taking into account the datasets of the various companies with the column regarding the price increase or decrease as dependent variable and all the other columns as independent variables.

By building these models it was possible to identify which were the most significant independent variables that could be taken into account for the next parts of the analysis.
In fact, through the logit models it was possible to observe that among the most significant variables there were the variable indicating whether a specific company was present or absent at a specific date in the Index in question, the values regarding the total return (gross dividends) with a 7-day lag, the average sentiment value of the news of the day and the number of trades of that day. It was therefore possible to notice that the fact that a company belongs to the Index in question at a specific time instant has an impact on the dynamics of prices of the stocks considered.

 In addition to this, also the average sentiment value of the news of the day as well as the number of trades of the stock also influence the fluctuations of prices in question. Moreover, it was also possible to observe that, among the variables with a 7-day lag, the total return (gross dividends) is a significant independent variable and thus also that the data concerning a specific week has an impact on the dynamic of prices of the following period. As it was stated earlier, these are the variables that were taken into consideration in order to implement the neural network.

The following part of the analysis, in fact, involved building the neural networks. To this regard, the aforementioned variables were taken into consideration as independent variables, while the dependent variable was the column indicating whether the specific company considered was present or absent at a specific date in the STOXX Europe 600 Index.

The entire timespan was thus divided into two parts, the first part corresponding to 80\% of the total data and the second part corresponding to 20\% of the total data; the former part was used to train the model while the latter was used to test the model in question.

As it was stated earlier, the length of the timespan depended on the availability of the data and therefore for some companies the model was built considering approximately 450 periods, for example in the case of Cancom, and for other companies the available periods were about 500 or more, as in the case of Rotork PLC.

\section{Outcome of data analysis}

As it was stated earlier, the first part of the analysis involved the employment of the logit model, through which it was possible to determine which are the most significant independent variables with regard to the analysis of the fluctuations of the prices; these variables were then selected in order to build the neural networks for each company considered. From the logit model it was thus possible to observe that among the most significant variables there is the independent variable indicating whether the company at the time instant considered is a constituent or is not part of the STOXX Europe 600 Index. In other words, the dichotomic independent variable which indicates with number 1 the fact that the company in question is part of the Index and with number 0 the fact that the company considered is not part of the STOXX Europe 600 Index is among the factors that determine the rise or the decline in the prices of the specific stock. Similarly, it was possible to observe that also the number of trades affects the dynamics of the fluctuations of the prices of the stock. In addition to this, the third independent variable that was identified as one of the variables that has an impact on the prices of the stock is the average sentiment value of the news of the day, which basically records the sentiment which can be derived from the analysis of the information taken from the news. Finally, the only variable belonging to the group of the independent variables with a 7-day lag that was selected among the most significant variables is the total return (gross dividends). More specifically, in this case the data concerning the previous week was recorded and taken into consideration and therefore it was thus possible to observe that also factors belonging to a different week actually had an impact on the fluctuation of the following period of time.
These four factors, as it was mentioned earlier, were the four independent variables that were taken into account to implement the neural networks for the companies considered. Through the code written in Python it was possible to use 80\% of the data to train the model and the remaining 20\% to test the model in question. Therefore, this procedure was repeated for each company taken into consideration and the various levels of accuracy were thus determined. The levels of accuracy varied according to the specific company considered, but overall the models built for the various companies showed accuracy values of about 70\%, as it can be seen from Table 1 (for the sake of simplicity just 10 of the 50 companies are included in the following table).

\begin{table}[h!]
\begin{center}
	\begin{tabular}{ c | c }
		\hline
		{\bf Company Name} & {\bf Accuracy}  \\
		\hline  
		Cancom &	68,76\%  \\
		\hline 
		Dechra Pharmaceuticals &	70,24\% \\
		\hline 
		Logitech International& 	71,46\% \\
		\hline 
		Clariant &	69,49\%  \\
		\hline 
		Deutsche Wohnen	& 70,57\%  \\
		\hline 
		DS Smith &	68,23\%  \\
		\hline 
		Rotork &	69,65\%  \\
	    \hline 
		Roche Holding &	70,71\%  \\
		\hline 
		RSA Insurance Group	& 71,43\%  \\
		\hline 
		RELX &	70,14\% \\
		\hline 
	\end{tabular}
\end{center}
\caption{Values of accuracy}
\end{table}

The authors, as it is specified also in the following section, believe that there is still room for improvement, since the values of accuracy could possibly increase if the model, for example, is constructed by taking into consideration other combinations of significant independent variables.
Finally, the values of accuracy therefore provide a sufficiently satisfactory result, since the values in question are above 65\% and since they are consistent throughout the various companies that have been taken into consideration during the various parts of the analysis.

\section{Conclusion}

Through the thorough data analysis it was thus possible to observe and understand the main determinants of the dynamics of the price. The first part involved the careful and attentive data selection and cleaning processes. Through this procedure it was possible to select with pre-set criteria which companies had to be taken into consideration. In the following part of the analysis the most significant independent variables were determined through the employment of the logit model. These variables were then taken into account to create the neural networks, which, applied to the different companies considered, showed accuracy values of approximately the 70\%, thus providing a sufficiently satisfactory result.

Different types of future research could be related to this work. To this regard, in fact, future work could deal with the expansion of the databases considered, for example by taking into account more companies and longer periods of time (when possible).

In addition to this, the focus of future work could also be quantifying the effect that the inclusion or the exclusion from the STOXX Europe 600 Index has on the fluctuations of prices.

Moreover, future work could also include the application of these analysis and models to other indexes and therefore, in other words, to different economic contexts.

In conclusion, a secondary type of analysis could also include the assessment of the impact of these models on other aspects of these companies or eventually indexes considered and therefore, on a larger scale, on various economic and financial scenarios.

\end{document}